\documentclass[manuscript]{aastex}
\usepackage{tikz}
\usepackage{amsmath}

\usepackage{graphicx}

\usepackage{pdflscape}

\slugcomment{\textbf{The Astronomical Journal} }

\shorttitle{Double Ciffreo}
\shortauthors{Kim}

\begin{document}

%\title{Peculiar Morphology of Comet 108P/Ciffreo}
\title{Comet 108P/Ciffreo: The Blob}
%% Use \author, \affil, and the \and command to format
%% author and affiliation information.
%% Note that \email has replaced the old \authoremail command
%% from AASTeX v4.0. You can use \email to mark an email address
%% anywhere in the paper, not just in the front matter.
%% As in the title, use \\ to force line breaks.

\author{
Yoonyoung Kim$^1$, David Jewitt$^{2}$, Jane Luu$^{3}$, Jing Li$^{2}$ and Max Mutchler$^4$
} 
\affil{$^1$Institute for Geophysics and Extraterrestrial Physics, TU Braunschweig, D-38106 Braunschweig, Germany}
\affil{$^2$Department of Earth, Planetary and Space Sciences, UCLA, Los Angeles, CA 90095-1567}
\affil{$^3$ Centre for Earth Evolution and Dynamics, University of Oslo, NO-0315 Oslo, Norway}
\affil{$^4$ Space Telescope Science Institute, Baltimore, MD 21218}

\email{yoonyoung.kim@tu-bs.de}

\begin{abstract}

Short-period comet 108P/Ciffreo is known for its peculiar double morphology, in which the nucleus is accompanied by a co-moving, detached, diffuse ``blob''.  We report new observations of 108P/Ciffreo taken with the Hubble Space Telescope and the Nordic Optical Telescope and use them to determine the cause of this unusual morphology.  The separation and the longevity of the blob across several orbits together rule out the possibility of a single, slow-moving secondary object near the primary nucleus.  We use a model of coma particle dynamics under the action of solar gravity and radiation pressure to show that the blob is an artifact of the turn-around of particles ejected sunward and repelled by sunlight.  Numerical experiments limit the range of directions which can reproduce the morphology and explain why the co-moving blob appearance is rare.

\end{abstract}


\section{INTRODUCTION}
\label{intro}
Short-period comet 108P/Ciffreo was discovered on UT 1985 November 8 (by J. Ciffreo, reported in \cite{Heudier85}). 
It is a Jupiter family comet (i.e.~a likely escapee from the Kuiper Belt), with orbital semimajor axis $a$ = 3.646 au, eccentricity $e$ = 0.579, inclination $i$ =  14.0\degr, corresponding to Tisserand parameter with respect to Jupiter $T_J$ = 2.711.   The orbital period is 7.0 years; the most recent perihelion occurred on  UT 2021 October 16 at $q$ = 1.660 au.   

A peculiar morphology was   reported  in late 2021, with the emerging appearance of a double object, consisting of a centrally condensed nucleus with a co-moving diffuse structure (the ``blob'') located to its east.  108P was also noted for its peculiar optical morphology in earlier apparitions.  For example, observers in 1985 December noted a ``detached dust tail'' \citep{Larson86,Levy86}, persisting to at least 1986 March (Chen and Jewitt 1994).   Archived images taken 2014 November - December\footnote{\url{http://www.aerith.net/comet/catalog/search.cgi}} \footnote{\url{http://www.astrosurf.com/cometas-obs/}} \footnote{\url{https://tinyurl.com/dzx5dabu}}, although obtained with small telescopes and of variable quality, show a morphology soon after perihelion similar to that in early 2022, while a single post-perihelion image from 2007 July does not.  In other comets, a double appearance normally indicates a nearly co-moving, secondary source displaced from, and independent of, the outgassing activity from the main nucleus (e.g. Boehnhardt 2004).  However, this interpretation raises the question of how the double morphology could persist in successive orbits. An alternative possibility is that the detached appearance is caused by projection of dust particles in a sunward fan and then turned around by solar radiation pressure \citep{Manzini22}. This interpretation is also puzzling since it raises the question of why it would apply only to 108P out of the hundreds of comets that have been imaged in the modern era.  

We obtained Hubble Space Telescope observations in order to examine the unusual structure of 108P at high angular resolution, with the particular objective of being to understand the nature and origin of its double appearance.   In addition, we secured images with the Nordic Optical Telescope in order to monitor the temporal development of the morphology of this unusual comet.  Our purposes in  this paper are to present the new observations and to suggest an interpretation.

\section{OBSERVATIONS}
Observations were obtained using the 2.5 m diameter Nordic Optical Telescope (NOT) located on La Palma, in the Canary Islands, with the telescope tracked to follow the non-sidereal motion of the comet.  We used the ALFOSC  (Andalucia Faint Object Spectrograph and Camera) imaging spectrometer camera at the f/11 Cassegrain focus.  ALFOSC uses a 2048$\times$2064 pixel ``e2v Technologies'' charge-coupled device (CCD), yielding an image scale 0.214\arcsec~pixel$^{-1}$ across a 6.5\arcmin$\times$6.5\arcmin~field of view.  Seeing was variable from night to night but stayed mostly within the range 1.0\arcsec~to 1.5\arcsec~FWHM (full width at half maximum).
 We used broadband Bessel B (central wavelength $\lambda_c$ = 4400\AA, FWHM $\Delta \lambda$ = 1000\AA), V ($\lambda_c$ =5300\AA, $\Delta \lambda$ = 800\AA), R ($\lambda_c$ = 6500\AA, $\Delta \lambda$ = 1300\AA) and the ``$i_{int} 797~157$'' ($\lambda_c$ = 7970\AA, $\Delta \lambda$ = 1570\AA) facility filters to measure the optical colors of 108P, obtaining a majority of the data in R with integration times of 150 s.  Flat fields for these filters were constructed from images of an illuminated spot inside the dome of the NOT, while the bias level of the CCD was measured from a set of images obtained in darkness at the end of each night.  Photometric calibration was obtained from observations of Landolt stars \citep{Landolt92} when the sky was clear and from field stars, using calibrations from the Gaia and Sloan DR14 digital sky surveys \citep{Blanton17}, otherwise.
 
Observations were secured using the 2.4 m diameter Hubble Space Telescope under program GO 16904 (three orbits).  We used the WFC3 camera, giving an image scale 0.04\arcsec~pixel$^{-1}$ over a 120\arcsec$\times$120\arcsec~field of view.  The  images were taken using the F350LP filter which has central wavelength $\lambda_c$ = 6230\AA, and a very large FWHM $\Delta \lambda$ = 4758\AA, providing maximum sensitivity on a solar spectrum source. In each HST orbit, we obtained five exposures of 360 s duration (1800 s per orbit).   We also examined archival HST images obtained  when $r_H$ = 2.750 au on UT 1999 August 15 under GO 8274 (PI: P. Lamy), but could not identify 108P within the field.

\begin{figure}
\epsscale{0.95}
\plotone{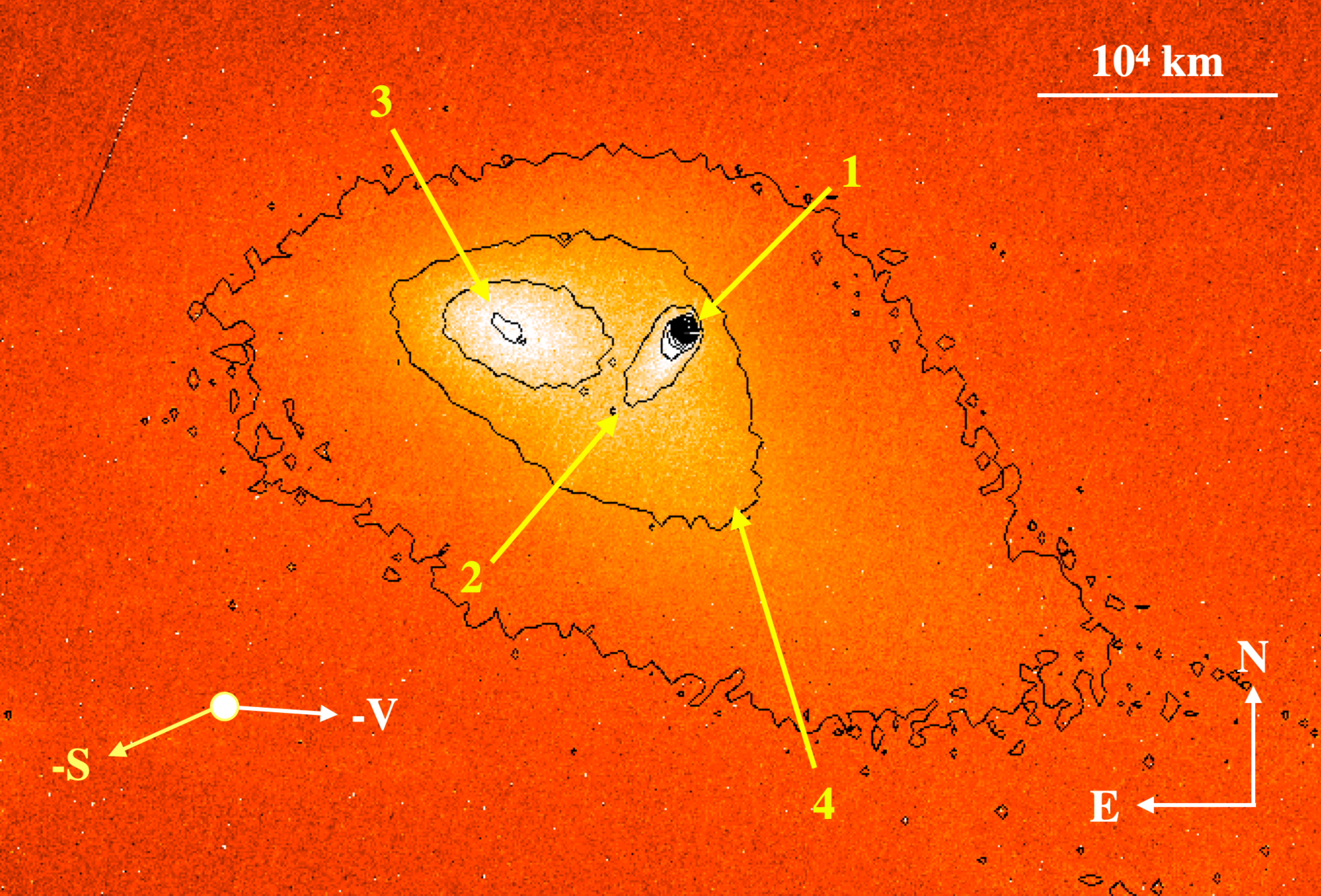}
\caption{HST image showing 108P on UT 2022 February 5 with isophotal contours overlaid.  A 10$^4$ km scale bar is shown, as well as the projected anti-solar ($-S$) and negative heliocentric velocity ($-V$) vectors. Numbered structures are 1) the central nucleus 2) jet 3) detached blob and 4) extended envelope.  \label{image_hst}}
\end{figure}

\begin{figure}
\epsscale{0.95}
\plotone{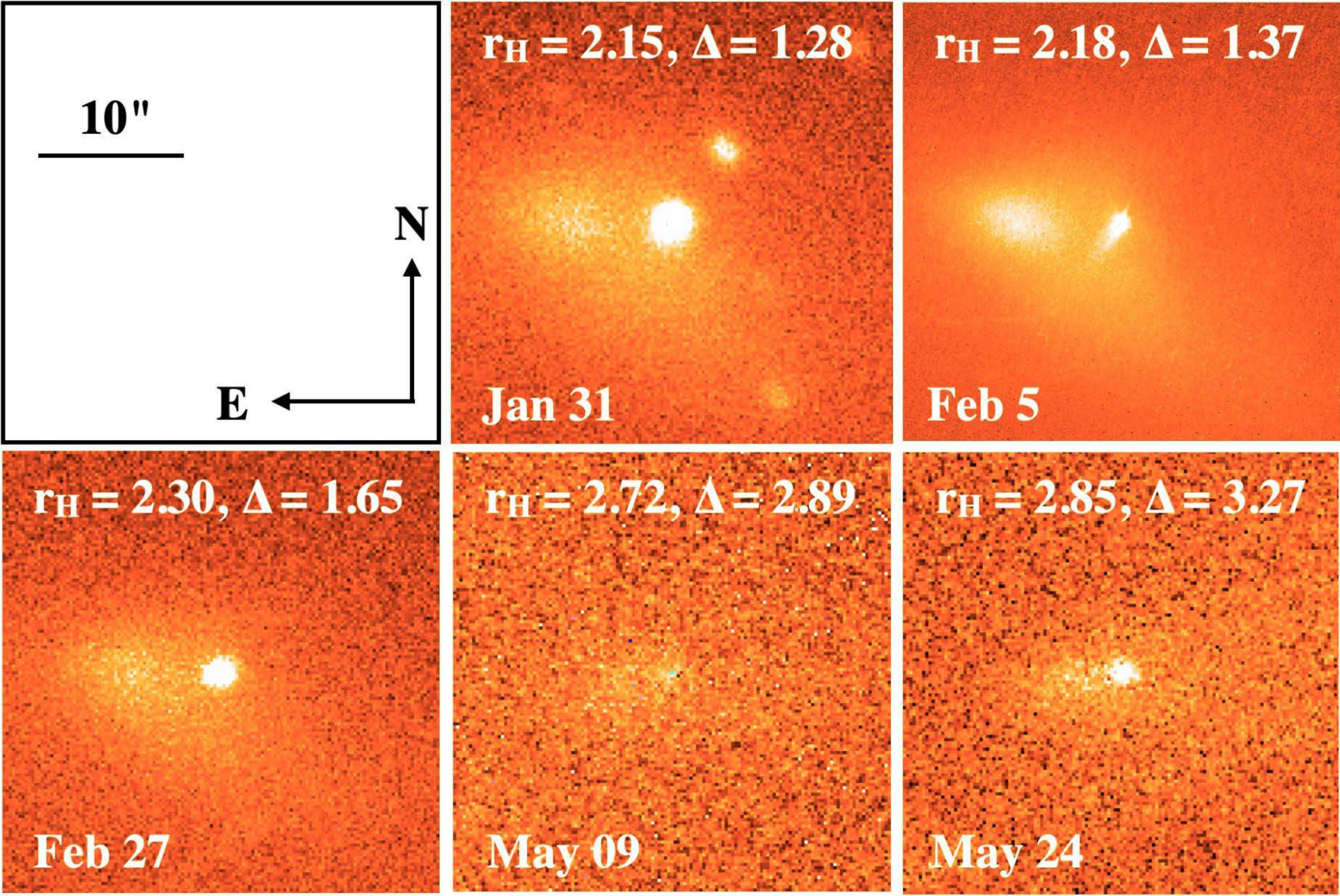}

\caption{Temporal development of 108P.  Each panel shows a region 30\arcsec$\times$30\arcsec~with North to the top and East to the left. The image from February 27 is the composite of B, V and R filter data. All others are R band images from the NOT except that of February 5, which shows an F350LP image from HST (HST images from February 10 and 13 are nearly identical to that from February 5 and are not separately shown).   
\label{composite}}
\end{figure}

The geometric circumstances of the observations are given in Table \ref{geometry}.  A median-combined image from HST is shown in Figure \ref{image_hst}, revealing  that 108P has four main components; (1) a centrally condensed bright core, (2) a linear, jet-like feature emanating from the nucleus, (3) a diffuse, approximately ellipsoidal blob projected $\sim$7\arcsec~(6580 km) to the east of the nucleus and  (4) diffuse material enveloping the other components.   The nucleus jet is not evident in images from the NOT, probably as a result of its lower angular resolution, but the other three components persist in all the observations taken with that telescope.   A composite of the images from different dates is shown in Figure \ref{composite}, where we show only one of the three HST visits (February 5) since the other two are taken close in time and are visually indistinguishable.

\begin{figure}
\epsscale{1.0}
\plotone{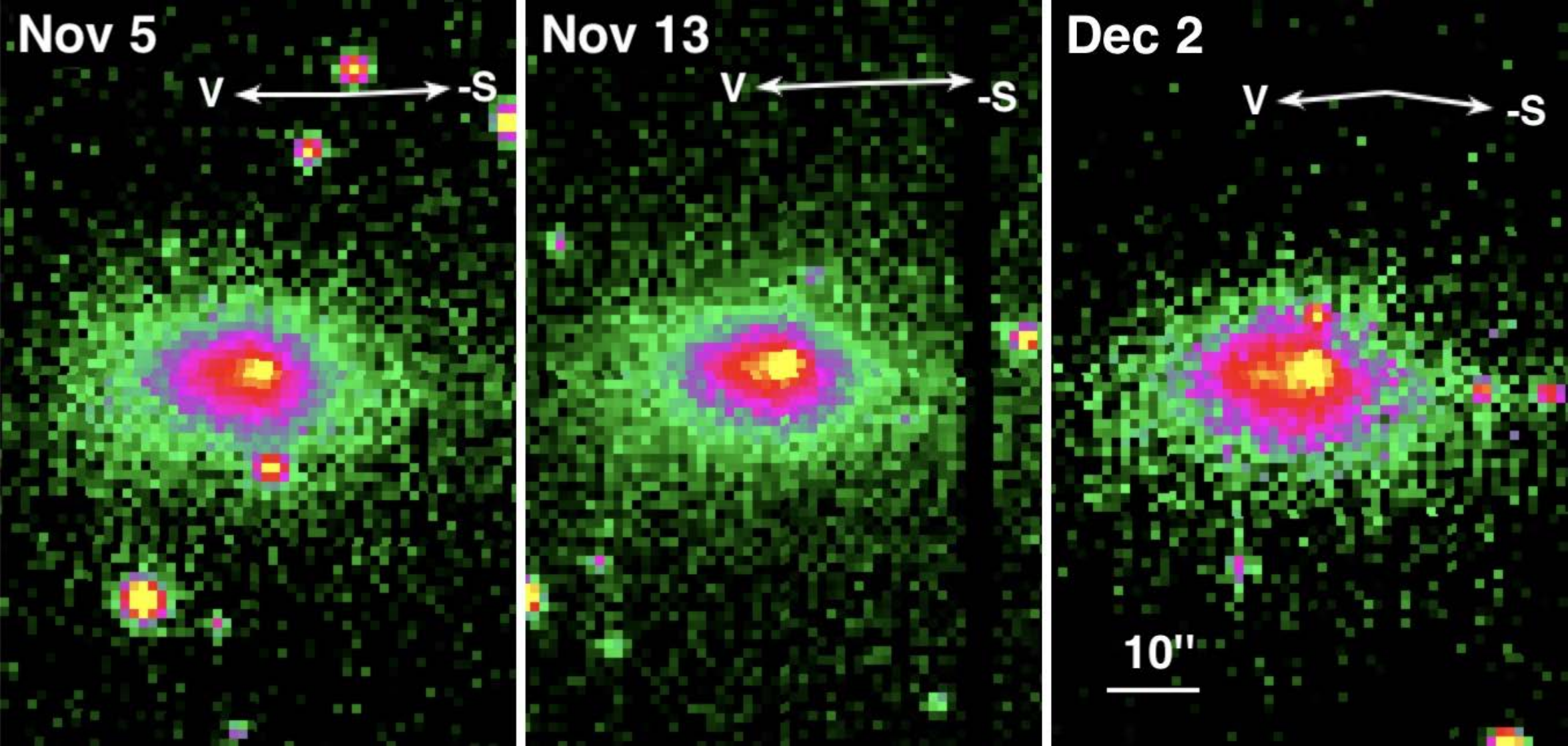}
\caption{Archival ZTF images of 108P taken on UT 2021 November-December, showing the emergence of the blob. A 10\arcsec~(7100 km) scale bar is shown, as well as the projected anti-solar ($-S$) and heliocentric velocity ($V$) vectors.
\label{ZTF}}
\end{figure}

We additionally identified and analyzed archival data from the ZTF (Zwicky Transient Facility; Bellm et al. 2019) taken  2021 November - December. These images have integration times of 30 s in the ZTF r-band filter. Although the ZTF images are of considerably poorer resolution (image scale 1\arcsec~pixel$^{-1}$),  the blob was nevertheless clearly apparent on UT 2021 December 2, and a slight extension in the nucleus was observed on UT 2021 November 13 (Figure \ref{ZTF}). Except for the emergence of the blob, the shape of the extended envelope in the background is stable, suggesting that the diffuse coma (component 4 in Figure \ref{image_hst}) is not related to the blob.

\textbf{Nucleus:} We used images from HST to set the strongest constraints on the nucleus.  Measurements obtained using a 0.2\arcsec~radius ($\sim$200 km) aperture, with sky subtraction from a surrounding annulus extending to 0.28\arcsec, are summarized in Table \ref{photometry}.  We corrected the apparent magnitude, $V$, to absolute magnitude, $H$ using phase coefficient $\beta$ = 0.02 magnitude degree$^{-1}$.  The absolute magnitude of the central region varies in the range 17.30 $\le H \le$ 18.08, but it is not clear if these variations reflect changes in the amount of near-nucleus dust, or changes in the projected cross-section of a rotating nucleus, or both.

The cross-section of the nucleus, $C_e$, is related to its absolute magnitude, $H$, and albedo, $p_V$, by

\begin{equation}
\label{eq1}
p_V C_e = 2.24\times10^{22} \pi \times 10^{0.4(V_{\odot} - H)}
\end{equation}

\noindent where $V_{\odot}$ = -26.73 is the absolute magnitude of the Sun \citep{Willmer18}.   Substituting for $H$ we find nucleus scattering cross-sections 0.9 $\le C_e(p_V/0.1) \le$ 1.8 km$^2$.   The equivalent circular radius  given by $r_n = (C_e/\pi)^{1/2}$, is 0.5 $\le r_n(p_V/0.1)^{1/2} \le$ 0.8 km.  Given the likelihood of dust contamination, however small, we take the smallest value $r_n = 0.5(0.1/p_V)^{1/2}$ km as our best estimate of the nucleus radius.  The gravitational escape speed from a spherical nucleus of density $\rho_n$ = 500 kg m$^{-3}$ and radius $r_n$ = 0.5 km is $V_e$ = 0.26 m s$^{-1}$.  Lamy et al.~(2004) cite a larger nucleus radius, $r_n$ = 1.4 km, from unpublished work by Scotti (1995).  Given the angular resolution offered by HST and its greater ability to isolate the nucleus from surrounding coma, we suspect that the latter estimate reflects larger dust contamination of the near-nucleus region.

%Tancredi et al.~(2000) discarded (and did not list) measurements of 108P from their list of reliably measured comets.  Later, Tancredi et al.~(2006) cite $H$ = 18.0 but with a large uncertainty estimated as $\pm$0.6 to $\pm$1.0 magnitudes.  Magnitude $H$ = 18 corresponds to a nucleus radius $r_n$ = 0.75 km for nominal geometric albedo 0.05, while the JPL Horizons web site lists $r_n$ = 1.5 km.

\begin{figure}
\epsscale{0.75}
\plotone{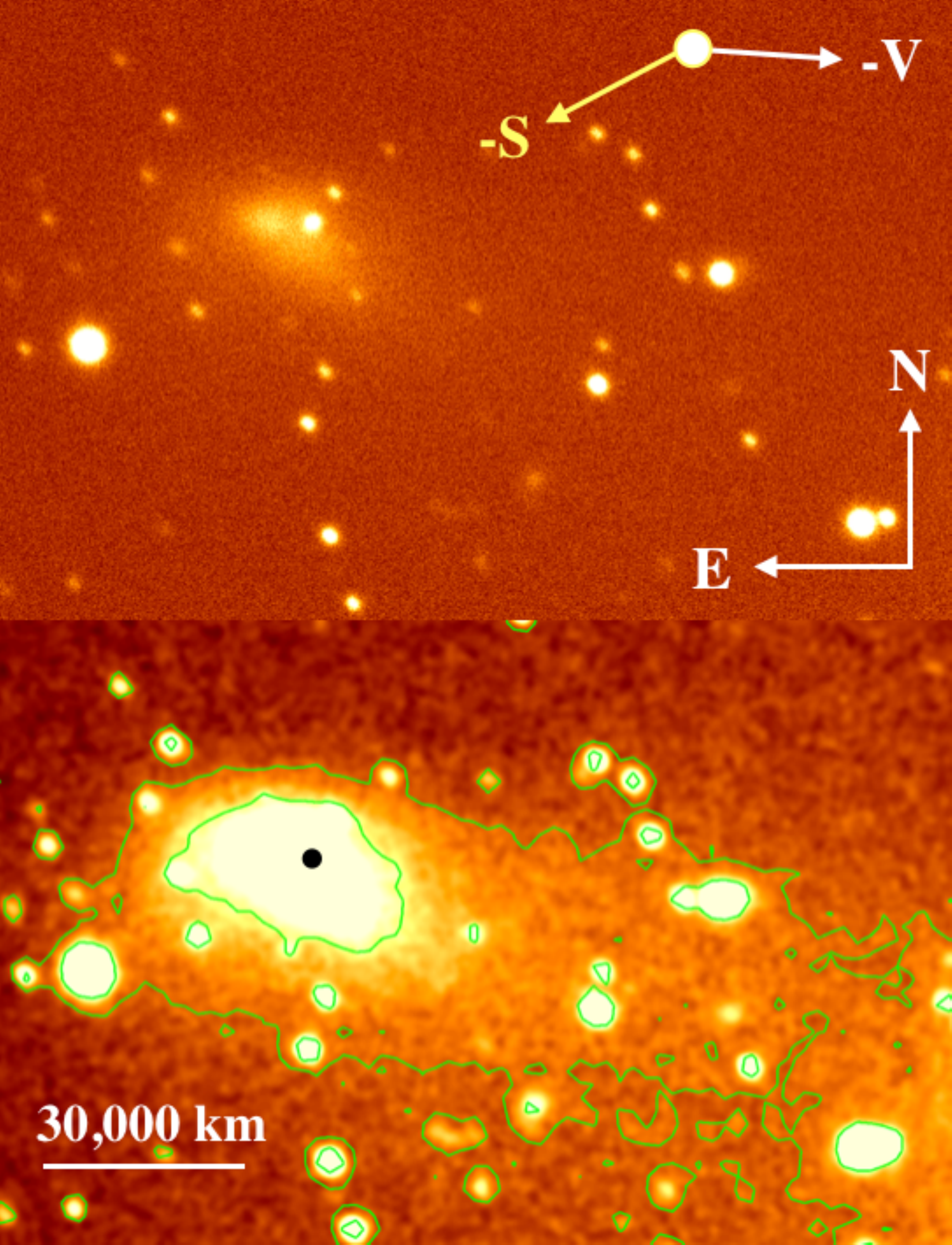}
\caption{(top) Composite of four 150 s exposures through the R filter taken on UT 2022 January 31 at the NOT.   (bottom) The same composite convolved with a 1.5\arcsec~FWHM Gaussian and contoured to show the tail of 108P.  Antisolar ($-S$) and negative heliocentric velocity vectors ($-V$) are shown, together with the cardinal directions and a scale bar. The filled black circle marks the location of the nucleus.  \label{tail}}
\end{figure}

\textbf{Diffuse Envelope:} 
A composite (600 s equivalent exposure) R-band image taken with the NOT from UT 2022 January 31 is shown in the upper panel of Figure \ref{tail}.   The same image, convolved with a Gaussian of 1.5\arcsec~FWHM and subject to a harsh stretch is shown in the lower panel, revealing a long, broad dust tail to the south-west (position angle $\sim$254$\pm$5\degr).  Viewed from out-of-plane angle $\delta_{\oplus}$ = -8.7\degr, the width of this tail is strongly affected by projection and cannot be used to directly determine the extent of the dust perpendicular to the orbit plane.  The surface brightness of the tail is shown  contoured, with the inner and outer contours corresponding to surface brightnesses $\Sigma$ = 24.3 and 25.4 magnitudes arcsecond$^{-1}$, respectively, each accurate to about $\pm$0.2 magnitudes arcsecond$^{-1}$ owing to sky brightness variations.  

We used large aperture photometry to estimate the total brightness of the comet, a measure of the sum of the cross-sections of the ejected dust particles.  Arbitrarily large photometry apertures are precluded by the influence of sky noise and background object contamination.  We present measurements of fixed apertures with projected radii 10$^4$ km and 1.5$\times10^4$ km.  The use of fixed linear apertures provides a measure of the light scattered by particles from within a fixed volume around the nucleus.  Note that some light from the coma and tail is unavoidably excluded even from the largest aperture.   The measurements are summarized in Table \ref{photometry}.  

\begin{figure}
\epsscale{1.0}
\plotone{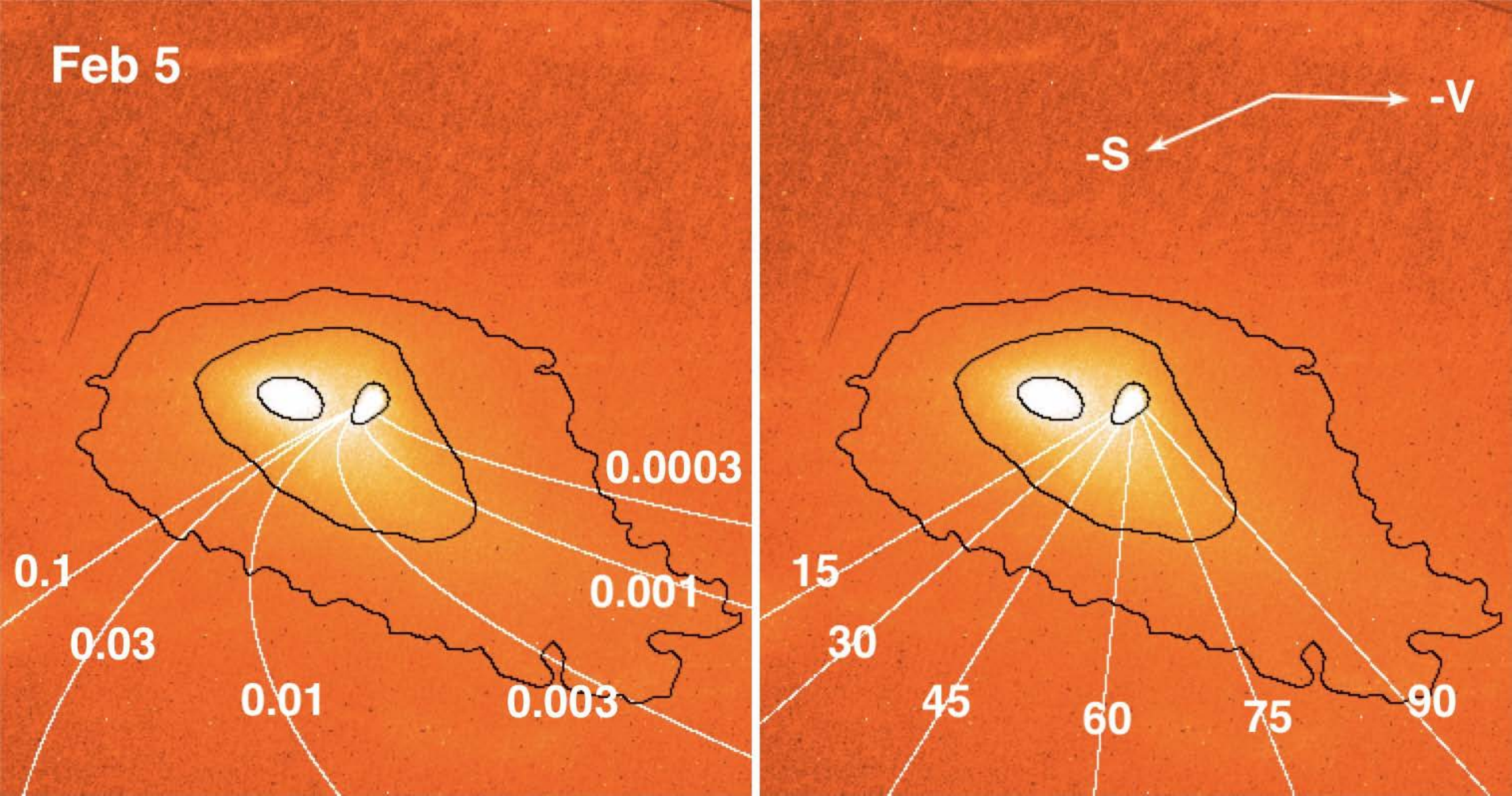}
\caption{(Left) syndynes for the UT 2022 February 5 image, showing the paths of particles with $\beta$ = 0.1, 0.03, 0.01, 0.003, 0.001, and 0.0003 and (right) synchrones computed for ejection 15, 30, 45, 60, 75, and 90 days prior to the date of observation.
\label{synsyn}}
\end{figure}

The motion of a cometary dust particle of radius, $a$, is controlled by $\beta$, the ratio of radiation pressure acceleration to solar gravity.  The radiation pressure acceleration on a solid, spherical, non-metallic grain is approximately given by $\beta = a_{\mu m}^{-1}$ (Bohren \& Huffman 1983).
Figure \ref{synsyn} (left) shows syndynes on UT 2022 February 5, which are positions of particles of a given $\beta$ released over a range of times (Finson \& Probstein 1968).  The direction of the extended envelope are matched by syndynes with $\beta \sim$ 0.0003 to 0.001, corresponding to particle radii $a \sim$ 1 to 3 mm.

\textbf{Linear Feature: }
The position angle of the jet-like linear feature varies in the HST data, from $\theta_\mathrm{PA}$ = 145\degr$\pm$2\degr~on UT 2022 February 5, to 141\degr$\pm$1\degr~on UT 2022 February 10 to 139\degr$\pm$2\degr~on UT 2022 February 13.  The linear appearance suggests that this is a synchrone, defined by an impulsive ejection of particles having a range of sizes, subsequently size-sorted by radiation pressure.  Figure \ref{synsyn} (right) shows synchrones on UT 2022 February 5. The measured angles indicate ejection on UT 2021 December 25 $\pm$ 2. The synchrones for the two other epochs yield the same date of ejection.
No photometric instability could be found near this date in the online archives.

\textbf{Colors: }We measured the colors of 108P on UT 2022 February 27 using the broadband BVRI filters of the Johnson-Cousins system  (Table \ref{colors}). We used circular photometry apertures with several different radii in order to assess spatial variations in the color.  Note that, even the smallest aperture is dominated by light scattered from coma dust, not from the nucleus. Calibration of the data was obtained using nearly simultaneous measurements of Landolt stars at similar airmass.  We assessed the photometric uncertainties by comparing measurements from three images taken in each filter.  These uncertainties are a few $\times$0.01 magnitude, except for the largest aperture where fluctuations in the sky background grow in importance, particularly for the B filter.  Within these uncertainties, we find no evidence for spatial variation of the optical colors of 108P (Table \ref{colors}).  As shown in the table, the colors are redder than those of the Sun \citep{Holmberg06}, but unremarkable when compared to the average colors of Jupiter family comets taken from \cite{Jewitt15}.

\section{DISCUSSION}

\subsection{Blob} 
\label{blob}
\begin{figure}
\epsscale{0.8}
\plotone{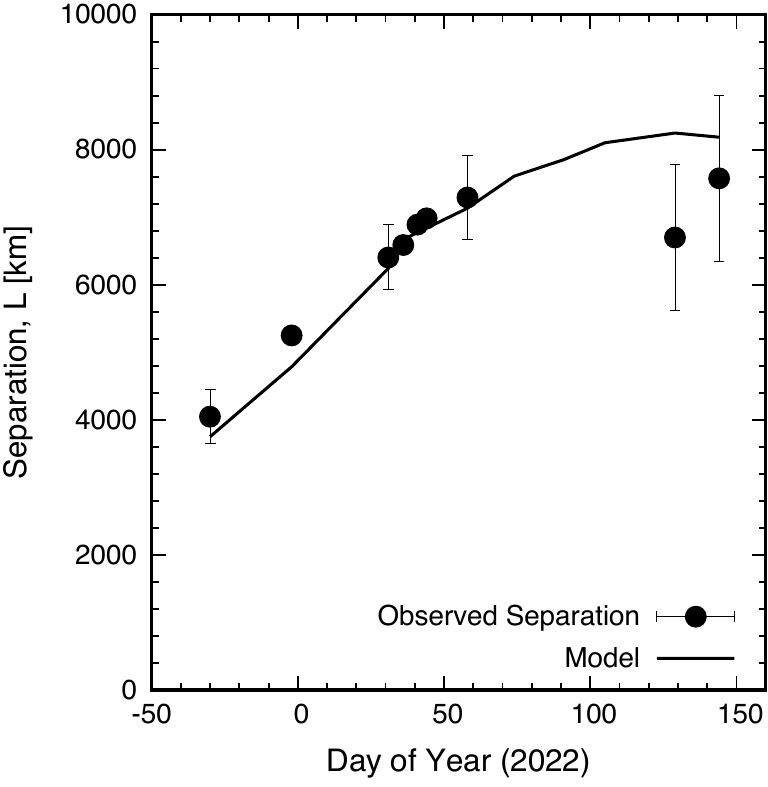}
\caption{Plane-of-sky separation between the nucleus and the blob as a function of the date of observation, expressed as Day of Year (DOY = 1 on UT 2022 January 1).  The solid line shows the time dependence of the separation computed from our model. 
\label{L_vs_DOY}}
\end{figure}

The comet shows remarkably little morphological development (Figure \ref{composite}), consistently exhibiting a bright nucleus region with a diffuse blob projected to its east.  Measurements of the separation between the nucleus and the centroid of the blob, determined visually, are listed in Table \ref{separation}, where the uncertainties correspond to $\pm$0.1\arcsec~in the HST observations and $\pm$0.5\arcsec~in the ground-based data.   
The linear separation, $L$ (km),  increases slowly over $\sim$6 months of observation (Figure \ref{L_vs_DOY}), reminiscent of the separation speeds of some fragmenting comets and asteroids \citep{Boehnhardt04}.
Extrapolation of the data gives $L = 0$ in 2021 August (DOY -140), 2 months before perihelion.
Figure \ref{L_vs_DOY} does not by itself exclude the possibility that the double appearance of 108P is caused by a splitting event.  On the other hand, the re-appearance of the double morphology with a $\sim$7\arcsec~separation in successive (seven year period) orbits argues strongly against this possibility.  %In the 7 year orbit period of 108P, the distance moved by a fragment at 0.44 m s$^{-1}$ would be 97,000 km, about 15 times the apparent separation.   
We further exclude the possibility that a fragment could be co-moving with the main nucleus because it is gravitationally bound. The $L \sim$6000 km separation is far larger than $r_{Hill}$, the Hill radius of the nucleus (which, for radius $r_n$ = 0.5 km and density $\rho_n$ = 500 kg m$^{-3}$ is $r_{Hill} \sim$ 100 km).

An alternative explanation is that the co-moving blob is a projection of the sunward fan into the plane of the sky \citep{Manzini22}.  Particles ejected sunward from the dayside of the nucleus experience radiation pressure from the Sun, of magnitude $\beta g_{\odot}$, where $\beta$ is a dimensionless number inversely related to the particle size and $g_{\odot}$ is the local solar gravitational acceleration.  We write $g_{\odot} = g(1)/r_H^2$, where $r_H$ is the heliocentric distance expressed in au and $g(1)$ = 0.006 m s$^{-2}$ is the acceleration at $r_H$ = 1  au. Given that the acceleration on a given particle is a constant, the turn-around distance representing the ``nose'' of the coma in the direction to the Sun is simply

\begin{equation}
L = \frac{U^2}{2\beta g_{\odot}}
\label{L}
\end{equation}

\noindent where $U$ is the sunward ejection velocity.  In gas drag acceleration, the particle ejection speed is a function of both the outflow speed in sublimated gas, $V_s$, and the particle size.    As a rough approximation, we take the thermal speed $V_s = (8 k T/\pi \mu m_H)^{1/2}$, where $T \sim 200$~K is the temperature of the sublimating surface, $\mu$ = 18 is the molecular weight of water, the dominant molecule, and $m_H$ is the mass of the hydrogen atom.  Substitution gives $V_s \sim$ 500 m s$^{-1}$. The smallest particles are well-coupled to the gas and leave the nucleus at approximately, $V_s$. Larger particles are accelerated to smaller terminal ejection speeds, given by $U = V_s a_{\mu m}^{-1/2}$, where $a_{\mu m}$ is the particle radius expressed in microns and $\beta = a_{\mu m}^{-1}$.  
Substituting for $U$ in Equation \ref{L} we have

\begin{equation}
L = \frac{4 k T}{\pi \mu m_H} \frac{r_H^2}{g(1)}
\end{equation}

\noindent showing that $L$  is independent of the particle size.   Substitution, taking $r_H$ = 2.2 au as the middle distance (Table \ref{geometry}) gives $L \sim 9\times 10^4$ km.  For comparison, the separation of the coma blob  from the nucleus is $\sim$7000 km in the plane of the sky and, corrected  for projection assuming that the blob lies on the Sun-comet line, is $L/\sin(\alpha) \sim$ 23,000 km.   Given the many approximations and assumptions, we regard this as reasonable agreement, consistent with the possibility that the separated blob is a projection of the sunward fan.

\subsection{Monte Carlo Dust Model}

To further explore the possibility that the blob is a turn-around projection of the sunward fan, we used a Monte-Carlo simulation of the dust. This model, developed by \citet{Ishiguro07} and used in \citet{Kim20}, accounts for the action of solar gravity and radiation pressure on particles drawn from a size distribution and released from the nucleus over a range of directions and speeds.  By its nature, the Monte-Carlo simulation cannot provide unique solutions for the dynamics and structure of the coma.  Its value lies in the identification of possible solutions while excluding combinations of parameters that are inconsistent with the data.

We consider a dust jet directed perpendicular to the surface of a rotating nucleus. The rotation of the nucleus sweeps the axis of the jet around a small circle centered on the pole (whose projected direction is ($\alpha_\mathrm{pol}$, $\delta_\mathrm{pol}$)) with a half-angle, $\omega$.
We assume that the effective time-averaged jet axis ($\alpha_\mathrm{jet}$, $\delta_\mathrm{jet}$) simply corresponds to the pole ($\alpha_\mathrm{pol}$, $\delta_\mathrm{pol}$) and that jet is only active when the source region on the nucleus is illuminated by the sun. The model assumes a power-law distribution of particle sizes, with differential index $q$ = 3.5, dust terminal ejection speed $V = V_0~\beta^{1/2}$ m s$^{-1}$, a dust production rate $\propto r_H^{-k}$, and a range of particle sizes from $\beta_\mathrm{min}$ to $\beta_\mathrm{max}$.
We assume that dust is ejected from $t_0$ to $t_1$, where $t_0$ is the time elapsed between the start of dust ejection and the observation, and $t_1$ is the time elapsed between the end of dust ejection and the observation. 
%dust emission onset in 2021 August (Section 3.1),
We created a number of model images using a range of parameters, and the model images were visually compared to the observations to find plausible parameters that give it a detached blob appearance.
%, and then we fitted the size, elongation, orientation, and the separation of the blob to find the best-fit parameters.
Numerical experiments limit the range of dust parameters:

\begin{itemize}

\item 
We find plausible solutions for jet direction of $190\degr\lesssim\alpha_{\rm jet}\lesssim210\degr$ and $0\degr\lesssim\delta_{\rm jet}\lesssim10\degr$ to produce a detached appearance and the observed orientation of the blob.
A collimated jet ($10\degr \lesssim \omega \lesssim 15\degr$) originating from the pole was required to fit the data.
%to produce a detached appearance and the observed elongation of the blob.
We find a best-fit pole orientation of $\alpha_{\rm pol}=$ 200$\degr$ and $\delta_{\rm pol}=$ 0$\degr$. The implied nucleus obliquity is $\varepsilon \sim 90\degr$ (i.e. the pole lies in the orbital plane).

\item
The largest particle size without destroying the detached blob appearance is given by $\beta_\mathrm{min} \sim 5\times10^{-5}$ (particle radii $\sim$20 mm).  %The upper limit to $\beta_\mathrm{max}$ is poorly defined.
The smallest particle size is poorly defined.
We set a lower limit to $\beta_\mathrm{max} > 5\times10^{-4}$ (particle radii $\lesssim$ 2~mm).  Smaller particles contribute less to the surface brightness of the blob because they are more quickly swept into the diffuse background.

% onset and termination
\item Dust ejection is assumed to begin in 2021 August (Section \ref{blob}). %The results are not strongly dependent on the assumed onset date. 
We find, by trial and error, that the termination point of the activity is 2021 December or earlier.
A later termination would require a steep dependence of dust production rate on heliocentric distance ($\propto r_H^{-9}$) for unambiguous separation of the blob, while moderate dependence ($\propto r_H^{-3}$) is found in other comets \citep{Ishiguro07}. The period of inferred activity to create a blob is $<5$ months.

\item
We find dust ejection speed $V = (115\pm5)~\beta^{1/2}$ m s$^{-1}$ best fits the observed separation of the blob. In Figure \ref{L_vs_DOY}, we show the best-fit modeled separations (solid line).
Particles with $2\le a \le20$~mm have speeds of ejection  in the range 0.8 $\le V \le$ 3 m s$^{-1}$.  Given this range of speeds, the travel time from the nucleus to the center of the blob lies in the range (4 to 13)$\times10^6$ s ($\sim$0.1 to 0.4 year), consistent with the six week ($\sim$0.1 year) lag between perihelion and the first appearance of the blob in 2021.

\end{itemize}

%Figure \ref{subsolar} shows the best solution for the projected pole direction (red circle) and the trajectories of the sunward directions projected $\pm$100 days from perihelion in the 1985 and 2021 perihelion passages, respectively. In the blob-emerging epochs in both 1985 and 2021, the pole points directly to the Sun at some point near perihelion. We additionally computed the subsolar colatitude (angle between the pole and the comet-Sun vector) of 108P as a function of time, expressed as DOY 2022. We found that the subsolar colatitude falls to zero at DOY = -145, consistent with the zero nucleus-blob separation in 2021 August (Section 3.1).

\begin{figure}
\epsscale{1.0}
\plotone{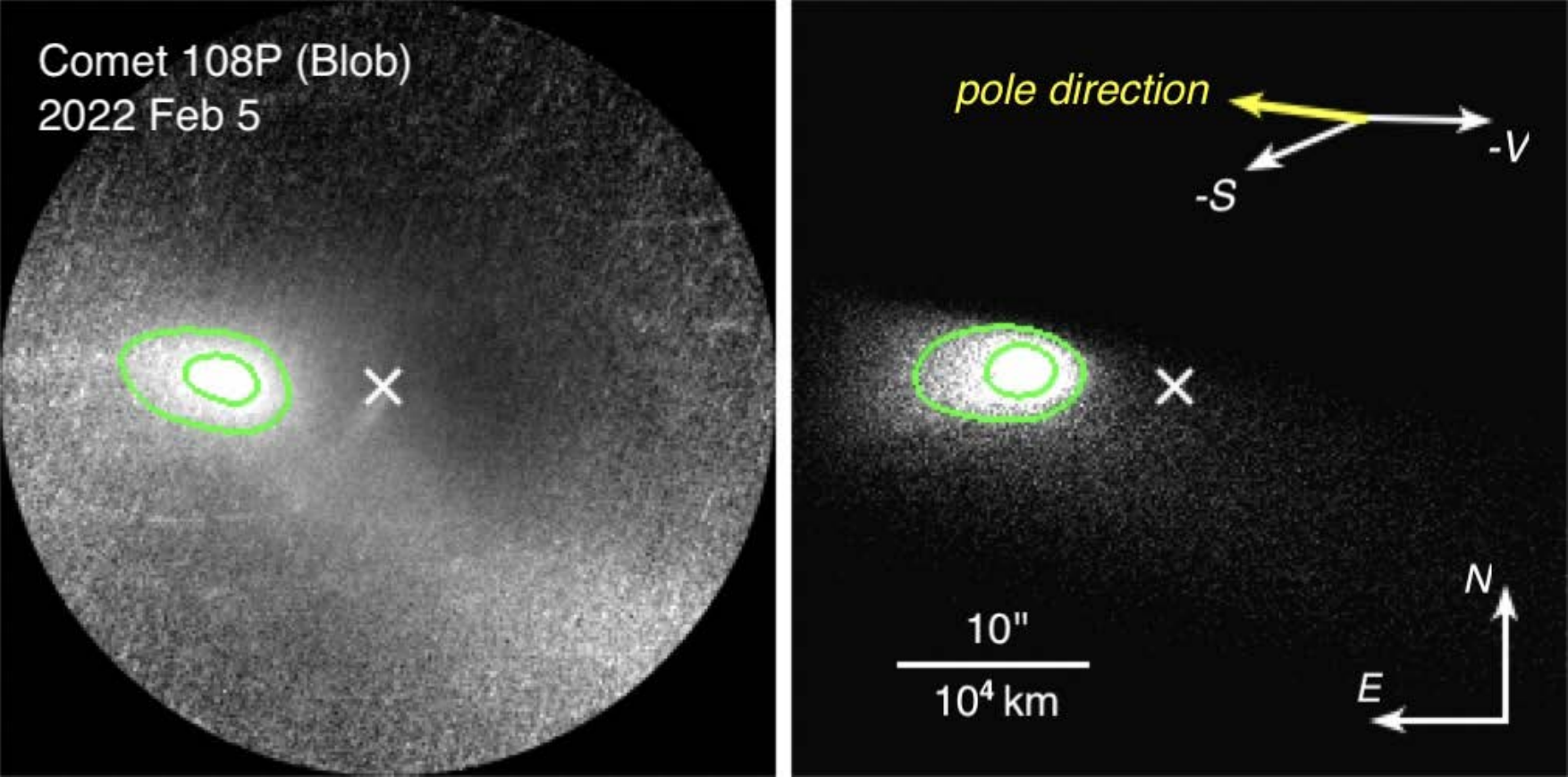}
\caption{(Left) UT 2022 February 5 image spatially filtered to suppress the coma by subtracting the median signal computed within a set of nucleus-centered nested annuli. (Right) best-fit Monte Carlo model on the same date. The nucleus location is marked with a cross in each panel. Isophotal contours are overlaid. The projected pole (i.e. jet) direction is shown, as well as the projected anti-solar ($-S$) and negative heliocentric velocity ($-V$) vectors. The model shows the turn-around of particles ejected from 2021 August to December.
\label{simulations}}
\end{figure}

\begin{figure}
\epsscale{0.85}
\plotone{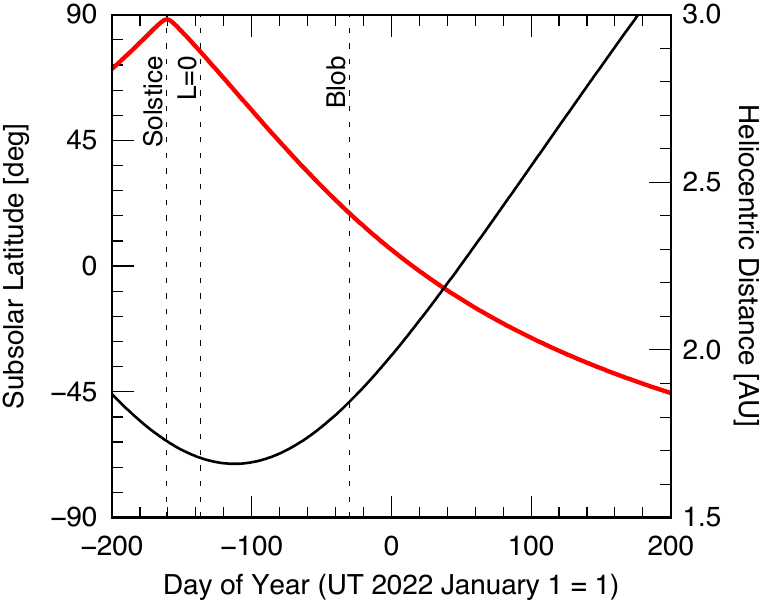}
\caption{Subsolar latitude (red line) of 108P as a function of time, together with the heliocentric distance (black line) on the right axis. We assumed a pole orientation of $\alpha_{\rm pol}=$ 200$\degr$ and $\delta_{\rm pol}=$ 0$\degr$ with an obliquity of 90$\degr$. Vertical lines indicate the dates of the northern solstice, inferred nucleus-blob separation $L=0$, and first detection of the blob.
\label{subsolar}}
\end{figure}

A specific example is given in Figure \ref{simulations}, with the best-fit parameters given in Table~\ref{parameter}.
In our model, the nucleus rotates with an obliquity of 90$\degr$, in which case the seasonal variation in solar insolation is maximum.
Figure \ref{subsolar} shows the subsolar latitude of 108P as a function of time. It is interesting to note that the northern solstice occurred in 2021 July, shortly before the inferred zero nucleus-blob separation in 2021 August.
%(i.e.~the onset time of dust jet).
This suggests that the emergence of the blob may be influenced by seasonal effects.

Lastly, we note that 108P must have multiple active regions, some of which may have formed an extended envelope (Figure \ref{image_hst}). The ZTF image sequence (Figure \ref{ZTF}) suggests that the detached blob and extended envelope have distinct origins. The focus of this paper is on the blob and modeling of the extended envelope is outside the scope of this paper.

\subsection{Pit Source}

Three features of the inferred jet are suggestive of a  pit source for the sublimated material, like the deep pits imaged on the nucleus of comet 67P \citep{Vincent15}.   
First, the required narrow collimation of the jet is consistent with self-shadowing by a topographic feature, as expected of sublimation from the bottom of a vertical pit.  Second, the apparently short-lived activity, or alternatively very steep heliocentric dependence of the production rate, is consistent with the limited period over which direct illumination of the floor of such a pit is possible. Third, the blob appeared just after the subsolar latitude 90$\degr$.

For a self-shadowed, rectangular pit source with a depth-to-diameter ratio $d/D$, the fraction of the floor illuminated by the Sun is

\begin{equation}
f(t) \sim 1-\frac{d}{D\tan (i(t))}
\label{pit}
\end{equation}

\noindent where the angle $i(t)$ is the Sun elevation from the pit floor \citep{Jewitt15b}. We assume that sublimation occurs at the pit floor. Equation \ref{pit} gives $f(t) \ge$ 0 for $i \ge i_c = \tan^{-1}(d/D)$,  and $f(t) = 0$ otherwise. For example, the floor of a pit source with $d/D=2$  will only be  half-illuminated when Sun elevation $i = 86\degr$ and in complete shadow when $i \le 63\degr$. If a deep pit source exists at the pole of comet 108P then a collimated jet would form in response to heating by the Sun when overhead. 
%\textbf{The maximum solar zenith angle to illuminate the floor, $\alpha_\mathrm{max}$, corresponds to the difference in true anomaly between Sun in zenith above pole and pit floor ($\tau_0$), and end of floor illumination ($\tau_1$), by $\alpha_\mathrm{max} = \tau_1-\tau_0$. 

We find that the floor of such a polar pit on 108P would be illuminated for about 60 days, broadly consistent with the longevity of the blob. Numerical experiments confirm that a collimated jet active for 60 days can generate a co-moving blob, whose longevity is further extended by the slow speeds and long travel times of the particles.

If such a pit source exists near the pole and the illumination conditions are satisfied, then similar blobs can be expected on other comets.  
We estimate the probability that a given nucleus would have an appropriately aligned pole as follows. Simulations  of 108P show that the pole direction must be in the range $190\degr\lesssim\alpha_{\rm pol}\lesssim210\degr$ and $0\degr\lesssim\delta_{\rm pol}\lesssim10\degr$.
The combination of $\Delta \alpha_{\rm pol} \sim 20\degr$ (0.35 radians) and
$\Delta \delta_{\rm pol} \sim 10\degr$ (0.17 radians) corresponds to $0.35\times0.17/2\pi \sim$0.9\% of the sunlit hemisphere.  On this basis, and assuming a random distribution of comet spin vectors, we would expect about 1\% of comets to show a double appearance like that of 108P due to a dust jet and particle turn-around.  However, this is an upper limit to the expected incidence of this distinctive morphology because not all comets necessarily possess sufficiently deep vents (indeed, these have only been clearly recorded on the nucleus of comet 67P) and because of the additional constraint from our models, namely that particles larger than $\sim$2 cm must be depleted or absent in order for the blob to be distinct.  Consistent with this expected low incidence rate, we find no clear examples of similar morphology in other comets in the refereed literature.  However, a nearly co-moving coma structure in the Centaur 174P/Echeclus (reported in Choi et al.~2006) is potentially of similar origin. This possibility deserves investigation.

\subsection{Mass}

We used aperture photometry to estimate the cross-section of the material in the blob.  For this purpose, we subtracted the median signal computed within concentric annuli centered on the main nucleus, using the on-line Cometary Coma Image Enhancement Facility software\footnote{Samarasinha, N., Martin, M., Larson, S., 2013. \url{https://tinyurl.com/yued8tva}}. The background-subtracted image from February 5 is shown in the left panel of Figure \ref{simulations}. In a circular aperture of projected radius 3.2\arcsec~($\sim$3200 km), with sky subtraction from a surrounding annulus extending to 6.4\arcsec, the apparent and absolute magnitudes are $V=24.4$ and $H=21.3$, respectively. The corresponding scattering cross-section computed from Equation (\ref{eq1}) is $C_e$ = 0.09 km$^2$.

The particle mass, $M_d$, and the scattering cross-section, $C_e$, are related by

\begin{equation}
M_d = \frac{4}{3} \rho \overline{a} C_e
\end{equation}

\noindent where $\rho$ = 500 kg m$^{-3}$ is the assumed particle density, $\overline{a}$ is the mean particle radius. We take $a_0$ = 1 mm and $a_1$ = 20 mm, yielding  $\overline{a} \sim$ 5 mm. Substituting $C_e$ = 0.09 km$^2$, we obtain $M_d = 3\times10^5$ kg for the mass in the blob particles, equivalent to an equal-density sphere  $\sim$5~m in radius. The ratio of the blob mass to nucleus mass ($r_n = 0.5$ km) is $M_d/M_n \sim 10^{-6}$.  This measurement shows how an almost inconsequential mass of material can substantially affect the morphology of the comet.

%\clearpage

\section{SUMMARY}

From new observations of short-period comet 108P/Ciffreo, we find the following.

\begin{itemize}
\item Comet 108P has a small nucleus with effective radius $\sim0.5(0.1/p_V)^{1/2}$ km, where $p_V$ is the unmeasured geometric albedo.  It exhibits a distinctive and recurrent double morphology not commonly reported in other comets.  
%The separation between the components is $L \sim$ 6000 km, about 20 times the Hill radius of a 1 km radius nucleus.  The upper limit to the rate of separation is set at $dL/dt \le$ 0.24 m s$^{-1}$.  

\item Recurrence of the double morphology in successive orbits excludes the possibility of a single, slow-moving secondary object near the nucleus of 108P.  

\item We used a Monte-Carlo model of coma particle dynamics to show that the blob is an artifact of the turn-around of particles ejected sunward in a narrow jet (half-angle 10\degr~$\lesssim \omega \lesssim$ 15\degr) and repelled by sunlight.  

\item We suggest that topography (likely a pit) present near the pole of a high-obliquity nucleus may be responsible for the collimation of the jet and the formation of the observed blob.

\item The blob morphology is reproduced only for a narrow  range of ejection directions, explaining why the appearance is uncommon.  We estimate that, if the spin vectors of cometary nuclei are randomly oriented,  $<$1\% of comets have the potential to exhibit such morphology.

\end{itemize}

\acknowledgments
We thank two anonymous reviewers and Jessica Agarwal for comments on the manuscript.
Based on observations made with the NASA/ESA Hubble Space Telescope, obtained from the data archive at the Space Telescope Science Institute. STScI is operated by the Association of Universities for Research in Astronomy, Inc. under NASA contract NAS 5-26555.  Support for this work was provided by NASA through grant number GO-16904 from the Space Telescope Science Institute, which is operated by AURA, Inc., under NASA contract NAS 5-26555. Y.K. acknowledges funding by the Volkswagen Foundation.

%% After the acknowledgments section, use the following syntax and the
%% \facility{} macro to list the keywords of facilities used in the research
%% for the paper.  Each keyword will be checked against the master list during
%% copy editing.  Individual instruments or configurations can be provided 
%% in parentheses, after the keyword, but they will not be verified.

{\it Facilities:}  \facility{HST, NOT, ZTF}.

%\clearpage

%% edition.

\begin{deluxetable}{lcccrrrrrrr}
\tabletypesize{\scriptsize}
%\rotate
\tablecaption{Observing Geometry 
\label{geometry}}
\tablewidth{0pt}
\tablehead{\colhead{UT Date \& Time} & \colhead{DOY\tablenotemark{a}} & \colhead{Tel\tablenotemark{b}}  & \colhead{Exp\tablenotemark{c}} & \colhead{$\nu$\tablenotemark{d}}  & \colhead{$r_H$\tablenotemark{e}} & \colhead{$\Delta$\tablenotemark{f}}  & \colhead{$\alpha$\tablenotemark{g}} & \colhead{$\theta_{- \odot}$\tablenotemark{h}} & \colhead{$\theta_{-V}$\tablenotemark{i}} & \colhead{$\delta_{\oplus}$\tablenotemark{j}}   }

\startdata
2021 Dec 02 08:39-08:40 & -30 & ZTF & $r$ & 44.4 & 1.849 & 0.980 & 19.9 & 261.6 & 275.1 & -3.9\\
2022 Jan 31 21:08-21:17 & 31 & NOT & R       & 68.7 & 2.150 & 1.281 & 16.1 & 118.2 & 267.9 & -8.7 \\ 
2022 Feb 05 12:34-13:07 & 36 & HST & F350LP & 70.2 & 2.175 & 1.336 & 17.5 & 114.3 & 267.6 & -8.4 \\
2022 Feb 10 08:28-09:01 & 41 & HST & F350LP & 71.9 & 2.203 & 1.398 & 18.7 & 111.0 & 226.4 & -8.1 \\
2022 Feb 13 09:30-10:03 & 44 & HST & F350LP & 72.9 & 2.220 & 1.439 & 19.4 & 109.3 & 267.4 & -7.8 \\
2022 Feb 27 23:15-23:49 & 58 & NOT & BVRI & 73.4 & 2.304 & 1.650 & 22.0 & 103.7 & 267.9 & -6.5 \\
2022 May 09 20:54-21:13 & 129 & NOT & R      & 95.3 & 2.723 & 2.886 & 20.5 & 99.2 &279.0 & -0.1 \\
2022 May 24 21:25-22:01 & 144 & NOT & R & 99.8 & 2.850 & 3.267 & 17.4 & 99.3 & 282.8 & +1.2 \\
\enddata

%% Text for table notes should follow after the \enddata but before
%% the \end{deluxetable}. Make sure there is at least one \tablenotemark
%% in the table for each \tablenotetext.

\tablenotetext{a}{Day of Year, 1 = UT 2022 January 1}
\tablenotetext{b}{Telescope: ZTF = Zwicky Transient Facility, NOT = Nordic Optical Telescope, HST = Hubble Space Telescope}
\tablenotetext{c}{Filter employed}
\tablenotetext{d}{True anomaly, in degrees}
\tablenotetext{e}{Heliocentric distance, in au }
\tablenotetext{f}{Geocentric distance, in au }
\tablenotetext{g}{Phase angle, in degrees }
\tablenotetext{h}{Position angle of projected anti-solar direction, in degrees }
\tablenotetext{i}{Position angle of negative heliocentric velocity vector, in degrees}
\tablenotetext{j}{Angle from orbital plane, in degrees}

\end{deluxetable}

%%%%%%%%%%%%%%%%%%%%%%%%%%%%%%%%%%%%%%%%%%%%%%%%%%%%%%%%%%%%%%%%%%%
\clearpage

\begin{deluxetable}{lrcrrrrrrrr}
%\tabletypesize{\scriptsize}
%\rotate
\tablecaption{Aperture Photometry\tablenotemark{a}
\label{photometry}}
\tablewidth{0pt}
\tablehead{\colhead{Date}  & \colhead{DOY\tablenotemark{b}} & \colhead{Inner\tablenotemark{c}} & \colhead{Middle\tablenotemark{d}} & \colhead{Outer\tablenotemark{e}}    }

\startdata
January 31	& 31	& -- & 16.38/13.86/42.9 		 & 16.07/13.55/57.1                     \\
February 5  	& 36 & 19.97/17.30/1.80 &  17.34/14.68/20.2                         &   17.05/14.38/26.5                        \\
February 10 	& 41 & 20.14/17.33/1.76 &  17.47/14.65/20.6                        &      17.17/14.36/27.1                     \\
February 13   & 44 & 20.99/18.08/0.88 &   16.90/13.99/38.0                       &     16.53/13.62/53.3                      \\
February 27  & 58 & -- &    18.43/15.09/13.8                       &   18.06/14.72/19.4                        \\
May 9   	& 129 & -- &   19.84/14.95/15.7                        &    19.43/14.54/22.8                       \\
May 24 	& 144 & -- &  20.19/15.00/15.0                         &   19.82/14.63/21.1                        \\

\enddata

%% Text for table notes should follow after the \enddata but before
%% the \end{deluxetable}. Make sure there is at least one \tablenotemark
%% in the table for each \tablenotetext.

\tablenotetext{a}{~The apparent red magnitude, $V$, the absolute red magnitude, $H$, and the scattering cross-section, $C_e$ (km$^2$), in the order V/H/$C_e$.}
\tablenotetext{b}{~Day of Year, 1 = UT 2022 January 1.}
\tablenotetext{c}{~HST-only photometry with a 0.2\arcsec~($\sim$200 km at the comet) radius  aperture.}
\tablenotetext{d}{~Photometry using a fixed aperture with a projected radius 10$^4$ km.}
\tablenotetext{e}{~Photometry using a fixed aperture with a projected radius 1.5$\times10^4$ km.}

%I used beta = 0.02 and alb p= 0.1
\end{deluxetable}

%%%%%%%%%%%%%%%%%%%%%%%%%%%%%%%%%%%%%%%%%%%%%%%%%%%%%%%%%%%%%%%%%%%
\clearpage

\begin{deluxetable}{lccrrrrrrrr}
%\tabletypesize{\scriptsize}
%\rotate
\tablecaption{Optical Colors 
\label{colors}}
\tablewidth{0pt}
\tablehead{\colhead{Ap\tablenotemark{a}}  & \colhead{B-V} & \colhead{V-R}  & \colhead{R-I } & \colhead{B-R}  }

\startdata
2.1\arcsec~(2.6$\times10^3$ km) & 0.77$\pm$0.03	&	0.50$\pm$0.03	&	0.47$\pm$0.03	&	1.27$\pm$0.04  \\
3.2\arcsec~(3.8$\times10^3$ km) &  0.78$\pm$0.03 & 0.49$\pm$0.03 & 0.50$\pm$0.03  & 1.27$\pm$0.04 \\
%84.  & & &  & \\
12.5\arcsec~(15.0$\times10^3$ km)  & 0.72$\pm$0.10 & 0.47$\pm$0.03 & 0.43$\pm$0.03 & 1.19$\pm$0.10 \\
\hline
JFC\tablenotemark{b}  &  0.75$\pm$0.02 & 0.47$\pm$0.02 & 0.43$\pm$0.02 & 1.22$\pm$0.02 \\
Sun\tablenotemark{c}  & 0.64$\pm$0.02 & 0.35$\pm$0.01 & 0.33$\pm$0.01 & 0.99$\pm$0.02 \\

\enddata

%% Text for table notes should follow after the \enddata but before
%% the \end{deluxetable}. Make sure there is at least one \tablenotemark
%% in the table for each \tablenotetext.

\tablenotetext{a}{Aperture radius in arcsecond (km)}
\tablenotetext{b}{Average color of Jupiter Family Comets, from \cite{Jewitt15}}
\tablenotetext{c}{Color of the Sun, from \cite{Holmberg06} }

\end{deluxetable}

%%%%%%%%%%%%%%%%%%%%%%%%%%%%%%%%%%%%%%%%%%%%%%%%%%%%%%%%%%%%%%%%%%%%%%%
\clearpage

\begin{deluxetable}{lccrrrrrrrr}
%\tabletypesize{\scriptsize}
%\rotate
\tablecaption{Component Separation 
\label{separation}}
\tablewidth{0pt}
\tablehead{\colhead{UT Date} & \colhead{DOY\tablenotemark{a}} & \colhead{$\Delta$\tablenotemark{b}}  & \colhead{$\theta$ \tablenotemark{c}} & \colhead{$L$\tablenotemark{d}}     }

\startdata
December 02 &	-30	&	0.980	&	5.7	&	4051$\pm$400\\
December 30 &	-2	&	1.019	&	$\sim$7	&	$\sim$5250\tablenotemark{e}\\
January 31 &	31	&	1.281	&	6.9	&	6408$\pm$480 \\
February 05 &	36	&	1.336	&	6.8	&	6586$\pm$100 \\
February 10 &	41	&	1.398	&	6.8	&	6892$\pm$105 \\
February 13 &	44	&	1.439	&	6.7	&	6989$\pm$108 \\
February 27 &	58	&	1.650	&	6.1	&	7297$\pm$620  \\
May 09 & 129		&	2.886 & 	3.2 	& 6700$\pm$1080  \\
May 24 &	144		&	3.267	&	3.2	&	7580$\pm$1230 \\

\enddata

%% Text for table notes should follow after the \enddata but before
%% the \end{deluxetable}. Make sure there is at least one \tablenotemark
%% in the table for each \tablenotetext.

\tablenotetext{a}{Day of Year, 1 = UT 2022 January 1}
\tablenotetext{b}{Geocentric distance, au}
\tablenotetext{c}{Separation angle, arcsec}
\tablenotetext{d}{Plane of sky length, km}
\tablenotetext{e}{From \citet{Manzini22}}

\end{deluxetable}

%%%%%%%%%%%%%%%%%%%%%%%%%%%%%%%%%%%%%%%%%%%%%%%%%%%%%%%%%%%%%%%%%%%

\clearpage
\begin{deluxetable}{lcc}
  \tablecaption{Dust Model Parameters \label{parameter}}
  \tablewidth{0pt}
\tablehead{\colhead{Parameter} & \colhead{Input Values} & \colhead{Best-fit Values}}

\startdata
$u_1$ & 0.5 & Fixed\\
$q$ & 3.5 & Fixed\\
%$k$ & 0--12 with 3 interval & 3--9\\
$k$ & 3 & Fixed\\
$\beta_\mathrm{max}$ & $10^{-4}$ to $10^{-1}$ & $\gtrsim 5\times10^{-4}$\\
$\beta_\mathrm{min}$ & $10^{-5}$ to $10^{-3}$ & $5\times10^{-5}$\\
$t_0$ (days)\tablenotemark{a} & 170 & Fixed\\
$t_1$ (days)\tablenotemark{b} & 0–100 with 10 interval & $\gtrsim 30$\\
$V_0$ (m s$^{-1}$) & 60--150 with 10 interval & 110--120\\
$\omega$ ($\degr$) & 10--40 with 5 interval & 10--15\\
$\alpha_\mathrm{jet}$ ($\degr$) & 0--360 with 5 interval & 190--210 \\
$\delta_\mathrm{jet}$ ($\degr$) & -90 to 90 with 5 interval & 0--10 \\

 \enddata
 \tablenotetext{a}{Time elapsed between the start of dust ejection and the observation (UT 2022 February 5).}
 \tablenotetext{b}{Time elapsed between the end of dust ejection and the observation (UT 2022 February 5).}
\end{deluxetable}

\clearpage
%%%%%%%%%%%%%%%%%%%%%%%%%%%%%%%%%%%%%%%%%
%%%%%%%%%%%%%%%%%%%%%%%%%%%%%%%%%%%%%%%%%
%%%%%%%%%%%%%%%%%%%%%%%%%%%%%%%%%%%%%%%%%

\clearpage
%%%%%%%%%%%%%%%%%%%%%%%%%%%%%%%%%%%%%%%%%
%%%%%%%%%%%%%%%%%%%%%%%%%%%%%%%%%%%%%%%%%
%%%%%%%%%%%%%%%%%%%%%%%%%%%%%%%%%%%%%%%%%

\clearpage
%%%%%%%%%%%%%%%%%%%%%%%%%%%%%%%%%%%%%%%%%
%%%%%%%%%%%%%%%%%%%%%%%%%%%%%%%%%%%%%%%%%
%%%%%%%%%%%%%%%%%%%%%%%%%%%%%%%%%%%%%%%%%

\clearpage
%%%%%%%%%%%%%%%%%%%%%%%%%%%%%%%%%%%%%%%%%
%%%%%%%%%%%%%%%%%%%%%%%%%%%%%%%%%%%%%%%%%
%%%%%%%%%%%%%%%%%%%%%%%%%%%%%%%%%%%%%%%%%

\clearpage
%%%%%%%%%%%%%%%%%%%%%%%%%%%%%%%%%%%%%%%%%
%%%%%%%%%%%%%%%%%%%%%%%%%%%%%%%%%%%%%%%%%
%%%%%%%%%%%%%%%%%%%%%%%%%%%%%%%%%%%%%%%%%

\clearpage
%%%%%%%%%%%%%%%%%%%%%%%%%%%%%%%%%%%%%%%%%
%%%%%%%%%%%%%%%%%%%%%%%%%%%%%%%%%%%%%%%%%
%%%%%%%%%%%%%%%%%%%%%%%%%%%%%%%%%%%%%%%%%

\clearpage
%%%%%%%%%%%%%%%%%%%%%%%%%%%%%%%%%%%%%%%%%
%%%%%%%%%%%%%%%%%%%%%%%%%%%%%%%%%%%%%%%%%
%%%%%%%%%%%%%%%%%%%%%%%%%%%%%%%%%%%%%%%%%

\clearpage

\end{document}